\documentclass[runningheads,a4paper]{llncs}

\usepackage[english]{babel}

\usepackage{booktabs}

\usepackage[%
rm={oldstyle=false,proportional=true},%
sf={oldstyle=false,proportional=true},%
tt={oldstyle=false,proportional=true,variable=true},%
qt=false%
]{cfr-lm}
\usepackage{graphicx}

\usepackage{paralist}
\usepackage{graphicx}
\usepackage{subcaption}

\usepackage{cite}

\usepackage[T1]{fontenc}

\usepackage[math]{blindtext}

\usepackage{csquotes}

\usepackage{microtype}

\usepackage{url}
\makeatletter
\g@addto@macro{\UrlBreaks}{\UrlOrds}
\makeatother

\usepackage[table,xcdraw]{xcolor}

\usepackage[
bookmarks=false,
breaklinks=true,
colorlinks=true,
linkcolor=black,
citecolor=black,
urlcolor=black,
pdfpagelayout=SinglePage,
pdfstartview=Fit
]{hyperref}
\usepackage[all]{hypcap}

\usepackage{todonotes}

\usepackage{xspace}

\newcommand{\ie}{i.\,e.,\ }

\DeclareFontFamily{U}{MnSymbolC}{}
\DeclareSymbolFont{MnSyC}{U}{MnSymbolC}{m}{n}
\DeclareFontShape{U}{MnSymbolC}{m}{n}{
    <-6>  MnSymbolC5
   <6-7>  MnSymbolC6
   <7-8>  MnSymbolC7
   <8-9>  MnSymbolC8
   <9-10> MnSymbolC9
  <10-12> MnSymbolC10
  <12->   MnSymbolC12%
}{}
\DeclareMathSymbol{\powerset}{\mathord}{MnSyC}{180}

\usepackage{soul}

\usepackage[acronym]{glossaries} 
\newacronym{ast}{AST}{Abstract Syntax Tree}
\newacronym{ct}{CT}{Combinatorial Testing}
\newacronym{czt}{CZT}{Community Z Tools}
\newacronym{dap}{DAP}{Debug Adapter Protocol}
\newacronym{dbgp}{DBGP}{Common DeBugGer Protocol}
\newacronym{gui}{GUI}{Graphical User Interface}
\newacronym{ide}{IDE}{Integrated Development Environment}
\newacronym{lsp}{LSP}{Language Server Protocol}
\newacronym{poc}{PoC}{Proof of Concept}
\newacronym{pog}{POG}{Proof Obligation Generation}
\newacronym{po}{PO}{Proof Obligation}
\newacronym{slsp}{SLSP}{Specification Language Server Protocol}
\newacronym{vdm}{VDM}{Vienna Development Method}
\newacronym{vscode}{VS Code}{Visual Studio Code}
\newacronym{CSV}{CSV}{Comma Separated Values}
\newacronym{ISQ}{ISQ}{International System of Quantities}

\usepackage{chngcntr}

\usepackage[capitalise,nameinlink]{cleveref}
\crefname{section}{Sect.}{Sect.}
\Crefname{section}{Section}{Sections}
\crefname{lstlisting}{Listing}{Listing}
\Crefname{lstlisting}{Listing}{Listing}

\usepackage{listings} 
\usepackage{times}    

\lstdefinelanguage{json}{
    basicstyle=\ttfamily\small, 
    numbers=left,
    stepnumber=1,
    numbersep=8pt,
    breaklines=true,
    frame=single,
    xleftmargin=.11\textwidth, 
    xrightmargin=.11\textwidth
}

\usepackage{vdmlisting}

\begin{document}
\pdfgentounicode=1

\title{International System of Quantities library in VDM}

\author{Leo Freitas
}
\authorrunning{ }

\institute{School of Computing, Newcastle University, \\
\email{leo.freitas@newcastle.ac.uk}
}
			
\maketitle
\setcounter{footnote}{0} 
\begin{abstract}
    The International Systems of Quantities (ISQ) standard was published in 1960 as an effort to tame the wide diversity of measuments systems being developed across the world, such as the centimeter-gram-second versus the meter-kilogram-second for example. Such a standard is highly motivated by the potential of ``trivial'' (rather error-prone) mistakes in converting between incompatible units. There have been such accidents in space missions, medical devices, etc. Thus, rendering modelling or simulation experiments unusable or unsafe. We address this problem by providing a \textbf{SAFE}-ISQ VDM-library that is: \textbf{S}imple, \textbf{A}ccurate, \textbf{F}ast, and \textbf{E}ffective. It extends an ecosystem of other VDM mathematical toolkit extensions, which include a translation and proof environment for VDM in Isabelle\footnote{\url{https://github.com/leouk/VDM_Toolkit/}}.
\end{abstract}


\section{Introduction}\label{sec:intro}

The International System of Quantities (ISQ) standard\footnote{\url{https://en.wikipedia.org/wiki/International_System_of_Quantities}} (also known as SI\footnote{\url{https://en.wikipedia.org/wiki/International_System_of_Units}}) represents the world's most widely used system of measurement. It has official status in nearly every country in the world, and is employed in a wide range of fields (\textit{e.g.}~science, technology, medicine, commerce, avionics, everyday use, \textit{etc.}). It is fundamental for both physics and engineering~\cite{bipm-jcgm:2012:VIM}. An interesting compendium of application and their relevance in various domains can be found in~\cite{NPL-book}. 

The standard comprises an elegant, minimal and coherent system of units of measurement, with the following seven base unit dimensions of length (in metres, \textbf{m}), mass (in kilograms, \textbf{kg}), time (in seconds, \textbf{s}), electric current (in amperes, \textbf{A}), thermodynamic temperature (in kelvins, \textbf{K}), amount of substance (in mols, \textbf{mol}) and luminous intensity (in candelas, \textbf{cd}). From these base units, the system can accommodate for an unlimited number of aditional (combined and converted) quantities (see~\Cref{tbl:SI}). 

Unit combinations are known as coherent derived units (or measurement systems) and can be represented as products of powers of other units. For instance, one can combine length and time to create units of velocity (\(\frac{m}{s}\)) and acceleration (\(\frac{m}{s^2}\));~and one can convert between units to create alternative interpretations of units like velocity in kilometers or miles per hour. Importantly, these combinations give rise to new units, and conversions give rise to bijective unit correspondences, unless conversions involve conversion schemas leading to approximation issues. 

Over twenty such coherent derived units are standard, named and routinely used. Further new (unamed) units can still be created (\textit{e.g.}~velocity as centimeter per year). This expressivity gives rise to multiple prefixes (\textit{e.g.}~kilo, centi, \textit{etc.}) and industry-standard conversion and combination schemes (\textit{e.g.}~\(\frac{km}{h}\) in \(\frac{miles}{h}\)). Thus, representing those notions formally is particularly important. 

As an example, Functional Mock-up Interfaces (FMI), which are used by various critical industries, have such notions for ISQ units\footnote{\url{https://fmi-standard.org/docs/3.0/\#_physical_units}}. That community would benefit from having ISQ units formalised.         

The Vienna Development Method (VDM) has has been widely used both in industrial contexts and academic ones covering several domains, such as of Security~\cite{Kulik&20,Kulik&21a}, Fault-Tolerance~\cite{Nilsson&18}, Medical Devices~\cite{Macedo&08}, \textit{etc}. We extend VDM specification support with a suite of tools and mathematical libraries\footnote{\url{https://github.com/leouk/VDM_Toolkit/}}. The work is also integrated within the VDM Visual Studio Code (VSCode) IDE\@.

We were inspired to develop the \texttt{ISQ.vdmsl} library due to the frustration of repetead error-prone (and tediuos) calculations involving unit use and conversions. The \texttt{ISQ} VDM library has been used in various industrial applications, such as embedded control systems, medical devices, chemical quantities composition (in drug dosage computation), complex scheduling time-scales, \textit{etc}.     

In this paper, we report on the recent extension of the VDM toolkit to support the ISQ standard, various conversion schemes, approximate precision, all base and derived representations, \textit{etc.}. We illustrate its use with a variety of scenarios frequently seen in practical applications of ISQ\@.    

\section{Background}~\label{sec:background}

ISQ has conventions corresponding to measurement-quantity category, abbreviations, dimensions and common names listed in~\Cref{tbl:SI}. Further background on ISQ itself is given in context of their corresponding definitions within the VDM library description.  
\begin{table}[htbp]
    \centering
    \begin{tabular}{lccll}
        \toprule 
        \textbf{Quantity}         & \textbf{Symbol}& \textbf{Dimension}  & \textbf{SI name} & \textbf{SI Symbol}  \\ \hline 
        length                    & \(l\)          & \(L\)        &  metre	   & \textbf{m}   \\ \midrule
        mass                      & \(m\)	       & \(M\)	      &  kilogram  & \textbf{kg}  \\ \midrule
        time                      & \(t\)          & \(T\)        &  second    & \textbf{s}   \\ \midrule 
        electric current          & \(I\)	       & \(I\)	      &  ampere    & \textbf{A}	  \\ \midrule 
        thermodynamic temperature & \(T\)	       & \(\Theta\)   &  kelvin    & \textbf{K}	  \\ \midrule
        amount of substance       & \(n\)          & \(N\)        &  mole      & \textbf{mol} \\ \midrule
        luminous intensity        & \(I_v\)	       & \(J\)	      &  candela   & \textbf{cd}  \\
        \bottomrule
    \end{tabular}
    \caption{Standard measurement system (SI) base quantities of measurement}\label{tbl:SI}
\end{table} 

ISQ unit conversions and combinations are notorioulsy error-prone, with well known examples of tragic accidents as a result of miscalculations\footnote{\url{https://www.simscale.com/blog/nasa-mars-climate-orbiter-metric/}}. For critical applications that rely on such unit conversions and combinations, this is particularly problematic. This motivated the creation of a formal environment to capture ISQ clearly and concisely. For example, multiple conversion schemes are available, such as British Imperial System to the stadard one (SI).       

The \texttt{ISQ} VDM library is heavily inspired by similar work provided for the Isabelle/HOL theorem prover~\cite{Physical_Quantities-AFP}. We integrate this library within the VDM VSCode extension\footnote{\url{https://marketplace.visualstudio.com/items?itemName=overturetool.vdm-vscode}}, as well as VDMJ~\cite{Battle09}.


\section{Design Principles}\label{sec:principles}

Our VDM \texttt{ISQ} library has four core design principles:

	\begin{enumerate}
		\item \textbf{S}imple:~ease of use is crucial, given ISQ unit combination and conversion are pervasive and rather menial, albeit error prone, task;
		 
		\item \textbf{A}ccurate:~ISQ conversion errors account for a considerable amount of modelling process inaccuracies; hence, we wanted a solution where multiple units and correspondences were possible;
		 
		\item \textbf{F}ast:~ISQ conversions can be numerous; hence, a computationally efficient solution is important;
		 
		\item \textbf{E}ffective:~there are multiple ISQ combinations possible; hence, practical use for a variety of such variations is important.   
	\end{enumerate}
        
These \textbf{SAFE} principles underpin the overall library design goals. Its architecture follows corresponding Isabelle/HOL ISQ structures~\cite{Physical_Quantities-AFP}. We also provide examples of more exotic measurement systems and conversion operators and order of magnitude-dependent approximation functions. Finally, we state some of the expected combination and conversion relationships as traces to be validated. These correspond in part to some of the theorems about ISQ properties as defined in the Isabelle/HOL ISQ library~\cite{Physical_Quantities-AFP}.   

\section{Library architecture}\label{sec:architecture}

Next, we present the library architecture. User models have to import \texttt{ISQ.vdmsl}, which provides various functionalities divided into six parts: 
\begin{enumerate}
    \item Dimensions (\textit{e.g.} length), dimension vectors (\textit{e.g.}~pressure as \(\frac{kg}{ms^s}\)), and operators;
    \item Quantities and operators (\textit{e.g.}~dimension amounts like \(20\frac{kg}{ms^s}\));
    \item Measurement systems and operators (\textit{e.g.}~group of dimensions like BIS or SI); 
    \item Base and derived units;
    \item Scaling and conversion over quantities and measurement systems;
    \item Common prefixes (\textit{e.g.}~alternative names); and 
    \item Non-decimal measurement systems (\textit{e.g.}~British Imperial System). 
\end{enumerate} 

 
\subsection*{ISQ dimensions and dimension vectors}

\begin{figure}[htbp]
    \centering
        \includegraphics[width=\textwidth,scale=0.4]{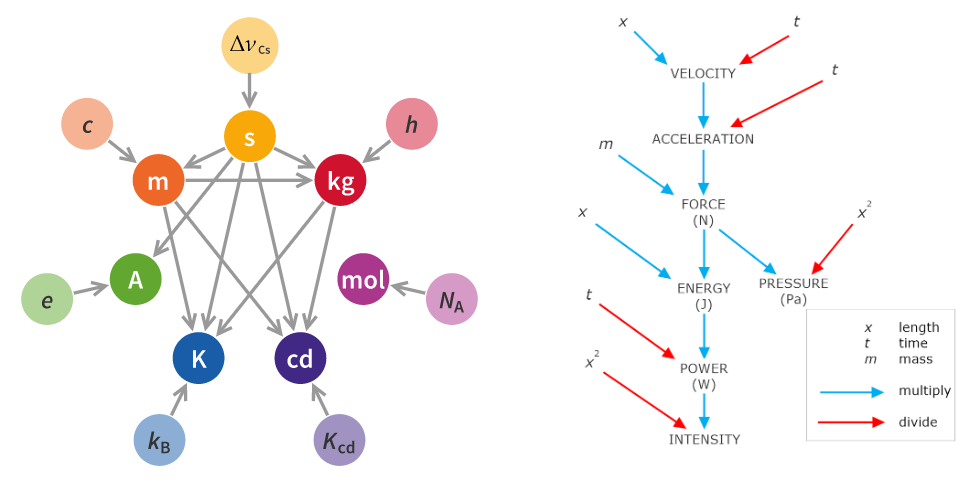}
    \caption{ISQ units of measurement and their conversion relationships.}\label{fig:UnitRelationships}
 \end{figure}

ISQ dimensions and dimension vectors represent base units of measurment and how they are combined and converted. There are seven base units of measurement are:~\textbf{L}ength, \textbf{M}ass, \textbf{T}ime, Electric Current (\textbf{I}), Temperature (\(\mathbf{\Theta}\)), Amount of Substance (\textbf{N}), and Luminous Intensity (\textbf{J}). Their standard quantities are, respectively:~meter (m), kilogram (kg), second (s), ampere (A), kelvin (K), mole (mol) and candela (cd). Their relationships and how they used for combination and conversions is illutrated in~\Cref{fig:UnitRelationships}\footnote{Figure originally appeared in~\url{https://en.wikipedia.org/wiki/International_System_of_Units}.}.  

Dimension vectors are maps capturing how each dimension relates. For instance, the notion of pressure can be measured in a derived unit named \textit{Pascal}, which is kilo per metre per second per second (\(\frac{kg}{ms^2}\)). 

These base unit combinations are captured in a so-called dimension vector, which is modelled as a VDM map, from dimension to an integer that captures the magnitude of each unit. In the \textit{Pascal} case, a single notion of mass (\(M\)) and corresponding inverted notions of length (\(\frac{1}{L}\)) and time (\(\frac{1}{T^2}\)) as (\(\frac{M}{LT^2}\)): 
\begin{vdmsl}[frame=none,basicstyle=\ttfamily\scriptsize]
    { <Length> |-> -1, <Mass> |-> 1, <Time> |-> -2, ... }
\end{vdmsl}   
\noindent This dimension mapping enables the representation of a wide variety of base and derived units with many dimensions. To make conversion operations (see below) homogeneous and compositional, dimension vectors must describe all base unit dimensions. In the example above, all remaining base units are mapped to zero.

There are \(22\) predefined dimension vectors:~seven for each of the base units mentioned; and fifteen for other so-called coherent derived units. These derived units combine one or more of the base units and are for:
%
%
area (\(L^2\)), volume (\(L^3\)), frequency (\(\frac{1}{T}\)), velocity (\(\frac{L}{T}\)), acceleration (\(\frac{L}{T^2}\)), energy (\(\frac{L^2M}{T^2}\)), power (\(\frac{L^2M}{T^3}\)), force (\(\frac{LM}{T^2}\)), pressure (\(\frac{M}{LT^2}\)), charge (\(IT\)), potential difference (\(\frac{L^2M}{T^3I}\)), capacitance (\(\frac{T^4I^2}{L^2M}\)), radian (\(\frac{L}{L}\)), steradian (\(\frac{L^2}{L^2}\)) and wattage (\(\frac{L^2M}{T^3}\)). 
Each of these derived units has a corresponding dimension vector as a VDM map constant. 

It is also possible to define dimensionless vectors for so-called bare or pure quantities, such as radians. Even though they are non-SI units, such quantities are useful for various applications. These are dimension vectors where all omitted dimensions map to zero. Single quantities like meter or seconds, have dimension vectors with all other dimensions mapped to zero, but the one dimension in question that is mapped to one.  

Dimension vector operators are used to combine these derived dimensions, as well as to create new user-defined dimentions. These are recursive map operations, which ensure quantity scaling and conversion will be sound. Moreover, to ensure sound dimension conversions, dimension vectors must be total on all seven base units.      

\subsection*{\texttt{ISQ} quantities}

ISQ quantities represent the magnitude of a specific dimension vector. Unit quantities (of magnitude \(1\)) for each of the base and derived units are used to compute conversions between measurement systems (\textit{e.g.}~kilometer per hour into miles per hour). For example, the unit quantity of area is defined as
\begin{vdmsl}[frame=none,basicstyle=\ttfamily\scriptsize]
    UNIT_AREA: Quantity = mk_Quantity(1, {<Length> |-> 2, ...});
\end{vdmsl}
\noindent where all other base units in the area dimension vector are mapped to zero. 

Quantities comparison (or ordering) are also limited to those within the same dimension vector, given that it does not make sense to compare units of area and velocity, for example:
\begin{vdmsl}[frame=none,basicstyle=\ttfamily\scriptsize]
    Quantity :: m: Magnitude d:- DimensionVector 
    ord q1 < q2 == q1.m < q2.m; 
\end{vdmsl}
\noindent A variety of other kinds of auxiliary quantities are defined, such as integer quantities, or single dimension quantities for each base and derived units. Notice that we use record equality abstraction for the dimension vector part of \verb'Quantity'. That is, the dimension vector map is not to be considered when comparing two quantities for equality.   

Operators between quantities include (but are not limited to): multiplication, replication, division, inversion, summation, (unary and binary) subtraction, scaling, \textit{etc}. They operate over the quantity's magnitude and perform any necessary dimension composition. Finaly, dimension composition operates in the same way as for quantities. For example, if users divide ten meters by two hours, they get the result in meters per hour. This is represented on the base units of the \texttt{SI} system, which is in fact just meter per second (the \texttt{SI} base units of length and time).    

\subsection*{\texttt{ISQ} measurement systems}

An ISQ measurement system represents a named quantity with a corresponding dimension vector conversion schema. A conversion schema defines (non-zero magnitude) factors to convert between other named measurement systems, for all base dimensions. 

Like dimension vectors, conversion schemas are VDM maps from all dimensions to a non-zero (real-valued) magnitude.    
They represent a mapping determining how each dimension magnitude is to be viewed within another measurement system. For example, the standard ISQ conversion schema (\texttt{SI}) is simply the identity map of all dimensions (\textit{e.g.}~one metre corresponds to one metre). More interestingly, the conversion schema for the British Imperial System (\textit{e.g.}~mile, yard, foot, \textit{etc.}) is named \texttt{BIS}. The measurement system representing yards in \texttt{BIS} is defined as     
\begin{vdmsl}[frame=none,basicstyle=\ttfamily\scriptsize]
  BIS  = { <Length> |-> 0.9143993, <Mass> |-> 0.453592338, <Temp> |-> 5/9, ... };
  YARD = mk_MeasurementSystem(mk_Quantity(1, LENGTH), BIS, "BIS");
\end{vdmsl}          
\noindent That is, a yard-unit quantity in \texttt{BIS} corresponds (will be converted into) \(0.9143993\) meters in the standard measurement system. The \texttt{LENGTH} dimension vector corresponds to dimension vector that maps all dimensions to zero and length to one.    

Crucialy, all ommited dimensions in \texttt{BIS} map to \(1\). This is different from dimension vectors that are mapped to zero by default. Such setup is crucial for how quantity dimensions are converted, which will be explained below.

With this representation, it is now possible to perform conversions between measurement systems through various operators. There are conversion schema operators for inverse, composion and scaling; and measurement system operators for replication, multiplication, inverse, division and various conversions between different measurement system's schemas. 

\subsection*{ISQ base and derived measurement systems}

For each of the base and derived dimensions and quantities, we define a corresponding measurement system. They are simple type instantations that represent the named abstractions to be used as types by end users. For example, the standard measurement system for area is defined as 
\begin{vdmsl}[frame=none,basicstyle=\ttfamily\scriptsize]
  Area = MeasurementSystem
  inv mk_MesurementSystem(mk_Quantity(magnitude, dimensions), schema, unit) == 
          magnitude = UNIT_AREA.dim and schema = SI and unit = SI_UNIT;
\end{vdmsl}
\noindent where \texttt{SI} is the conversion schema that maps all dimensions to \(1\), and \texttt{UNIT\_AREA} is the unit quantity for the dimension vector of area described above. Users can then extend this type to define their own notions of area within further refined (type invariants over) \texttt{Area} types.

These type extensions are performed for every base and derived dimentions. Furthermore, we define constants for the basic instance of such types (\textit{i.e.}~unit version of area within the \texttt{SI} standard measurement system). This in turn, enables the measurement system conversions described above, where the \texttt{SI} standard measurement system is the baseline.   

Further useful measurements are created to illustrate the flexibility of the framework. For instance, the functions below perfom the necessary conversions between \texttt{BIS} and \texttt{SI} measurement systems by scaling.   
\begin{vdmsl}[frame=none,basicstyle=\ttfamily\scriptsize]
    METRE: Metre = mk_MeasurementSystem(UNIT_LENGTH, SI, SI_UNIT);

    SI_YARD: () -> Metre
    SI_YARD() == scaleMS(0.9144, METRE);

    SI_FOOT: () -> Metre
    SI_FOOT() == scaleMS(1/3, SI_YARD());

    SI_MILE: () -> Metre
    SI_MILE() == scaleMS(1760, SI_YARD());
\end{vdmsl}
\noindent The first constant defines the baseline standard measurement system as a unit of length (\textit{i.e.}~a quantity with magnitude \(1\) and dimension vector mapping length to \(1\) and all other dimensions to \(0\)). The next functions transforms different imperial measurements by scaling their corresponding measurement system with their numerical magnitude correspondence (\textit{e.g.}~foot as third of a yard, yard as \(0.9144\) of a meter, \textit{etc.}). 

This kind of setup serves to illustrate how the various named abstractions can help qualify what computation is needed for what kind of conversion in as ituitive a fashion as possible, hence reducing room for error-prone conversion mistakes.  

\subsection*{ISQ scaling and conversion}

ISQ scaling is performed by multiplying the magnitude of the entities involved. For \texttt{Quantity}, that is direct multiplication of their magnitudes, whereas for measurement systems, it is the scaling of their corresponding quantities, defined as
\begin{vdmsl}[frame=none,basicstyle=\ttfamily\scriptsize]
    scaleMS: Magnitude * MeasurementSystem -> MeasurementSystem
    scaleMS(m, mk_MeasurementSystem(q, s, u)) == 
        mk_MeasurementSystem(scaleQ(m, q), s, u);

    scaleQ: Magnitude * Quantity -> Quantity
    scaleQ(m1, mk_Quantity(m2, d)) == mk_Quantity(m1 * m2, d);
\end{vdmsl} 
\noindent That is, measurement system scaling by magnitude \(m\) corresponds to scaling the measurement system's quantity magnitude by \(m\). 

Of course, when you scale two quantities, there is no change in their dimension vectors, whereas when you scale a measurement system, you must take dimension conversion into account. If you perform a measurement system operations over two different quantities' dimension vectors, then the result is in either a different dimension (\textit{e.g.}~if you divide quantities of length and time you get velocity) or in the dimension of one of the parameters (\textit{e.g.}~if you sum velocities in different units you must choose one as a resulting dimension vector). For example, if you perform a measurement system summation operation over two different quantities' dimension vectors, the result will be in the dimension of the leading operator (\textit{e.g.}~\(20\frac{miles}{hour} + 20\frac{km}{hour} = ?\frac{miles}{hour}\)).

To perform such operations over different measurement systems, a leading measurement system conversion schema has to be considered (\(\frac{miles}{hour}\)). For the \texttt{ISQ.vdmsl} library, that is the conversion schema of the first parameter in a measurement system operator. In the example above, the result would be in miles per hour. To achieve this, we must get the trailing measurement system (\(\frac{km}{hour}\)) converted into the leading (\(\frac{miles}{hour}\)) one, then perform quantity scaling.

Given a target conversion schema (\texttt{cs\_conv}) and a measurement system, we convert the quantity by converting the magnitude according to the target schema (\texttt{cs\_conv}), keeping the given schema (\texttt{cs}). Next, to quantity-convert in multiple dimensions, we must use the set-product (\texttt{prods\_r}) of the integer exponentiation between the corresponding conversion schema (\texttt{cs(i)}) and dimension vector (\texttt{dv(i)}) dimensions, for all base dimensions. The conversion function itself filters out any unexpected dimensions (\ie~those outside both conversion schema and dimension vector). Nevertheless, we also thought to add their dimension domain equivalence as a precondition to document the expected use of the function, despite the fact it would work without it.  
\begin{vdmsl}[frame=none,basicstyle=\ttfamily\scriptsize]
    quant_conv: ConversionSchema * DimensionVector -> MagnitudeN0
    quant_conv(cs, dv) ==
        prods_r({ cs(i)**dv(i) | i in set dom cs inter dom dv })
    pre
        dom cs = dom dv;

    ms_quant_conv: ConversionSchema * MeasurementSystem -> MeasurementSystem
    ms_quant_conv(cs_conv, mk_MeasurementSystem(mk_Quantity(m, d), s, u)) ==
        mk_MeasurementSystem(mk_Quantity(quant_conv(cs_conv, d) * m, d), s, u);
\end{vdmsl}
\noindent This is why the defaults for conversion schema and dimension version dimensions are \(1\) and \(0\), respectively. Those dimension vectors with zero dimension, will lead to \(1\) given the exponentiation; others dimensions will be multiplied accordingly to the conversion schema provided conversion mangnitude (\texttt{cs(i)}).

The resulting set of magnitudes is then multiplied together, now that their conversion has taken place. The function \texttt{quant\_conv} is at the heart of how the whole conversion process works:~it takes into account the conversion schema for all dimensions, ignoring zero dimensions, with the resulting product being the right magnitude in the right dimension vector.   

\paragraph*{Dimension vector correspondences.}~
As with other areas of mathematics, the unit signature of operators determine the kind of activity being performed. For example, Joule per Kelvin (\(\frac{J}{K}\)) is a unit of entropy or heat capacity. Furthermore, the analysis of dimension vectors determines that every coherent (base or derived) SI unit can be written as a unique product of powers of the base units constants. 

For example, the \textit{Joule} (\(J\)) is defined as the unit of energy as \(\frac{kgm^2}{s^2}\). That is, a Joule is equal to the amount of work done when a force displaces a mass (one \(kg\)) through a distance (one \(m\)) with particular acceleration (\(\frac{m}{s^2}\)). It can also be viewed as the amount of pressure (in \textit{Pascal}, \(Pa = \frac{kg}{ms^2}\)) through a volume (\(m^3\)). That is, \(Pa . m^3\), which is equal to \(\frac{kgm^3}{ms^2}\) and can be simplified to a \textit{Joule}. 

Finally, an important aspect of unit conversion is the guarantee that the target unit is adequate to the corresponding task intended. That is, a quanity unit type invariant will enforce that the intended computation landed on the intended unit --- whichever form its analogous dimensions may take (see~\Cref{sec:Examples}). 

\subsection*{ISQ common prefixes}

Given that magnitude, quantity and measurement systems are to be used seamlessly by the user (\textit{i.e.}~\(10\frac{km}{h}\) as the magnitude \(10\), quantity \(10\) of a particular dimension for velocity (\(\frac{L}{T}\)), or as the quantity within a measurement system of kilometers per hour), these types are known as a \texttt{Prefix}. 

Prefixes allow users to perform various operations over notions of magnitude, quantity of measurement systems without concern as to whether they are given a magnitude, a quantity or a measurement system parameter. In practice, this uses VDM union types to make the library ``natural'' to use rather than having multiple versions of the same computation for slightly different projections of various types involved. 

For example, the prefix for \texttt{kilo} is defined as the \(10^3\) magnitude of any given prefix. Then, the scaling of a magnitude for any kind of kilo can be defined as
\begin{vdmsl}[frame=none,basicstyle=\ttfamily\scriptsize]
    PREFIX_KILO   : MagnitudeN0 = (10**3);

    kilo: Prefix -> Prefix
    kilo(x) == scale_prefix(x, PREFIX_KILO);

    scale_prefix: Prefix * Magnitude -> Prefix
    scale_prefix(x, p) == 
        cases true:
            (is_Magnitude(x))         -> x * p,
            (is_Quantity(x))          -> scaleQ(p, x),
            (is_MeasurementSystem(x)) -> scaleMS(p, x)
        end;
\end{vdmsl}
\noindent Now the user can use expressions like \texttt{kilo(METRE)} or \texttt{kilo(10)} to represent \(10^3\) units of length in the \texttt{SI} measurement system (\textit{i.e.}~a kilometer), or \(10^4\) units of magnitude in whichever quantity or measurement system it gets used. These convenience functions hide from the user the need to perform operations at the level of magnitude, quantity or measurement systems separately. 

Specific (canonical \texttt{SI}) measurement systems can also be used for standard conversion. For example, the function \texttt{metrify} expects a given measurement system quantity, which will be converted to the standard metric quantity (\textit{e.g.}~convert miles per hour into meters per second) as
\begin{vdmsl}[frame=none,basicstyle=\ttfamily\scriptsize]
    metrify: MeasurementSystem -> Quantity
    metrify(ms) == ms_conv(ms, SI).quantity;

    mph2mps: Magnitude -> Magnitude
    mph2mps(mph) ==	mag(metrify(scaleMS(mph, BIS_MILE_PER_HOUR())));
\end{vdmsl}
\noindent The \texttt{BIS\_MILE\_PER\_HOUR} is a measurement system that divides the mile by the hour measurement systems. Overall, even though these computations are somewhat involved, the named abstractions help understanding and introspection of how various concepts are combined and converted.  

\subsection*{\texttt{ISQ} other (non-decimal) measurement systems}

We define various narrowing types and conversion functions between other (non-decimal) measurement systems. Notably, the one for the British Imperial System (\textit{i.e.}~yard, mile, \textit{etc.}) and the one for standard date and time (\textit{i.e.}~minute, hour, week, \textit{etc.}). 

Users can easily follow the examples to explore other exotic measurement systems they might be interested in representing, where the combination and conversion operators will be straightfoward (and hopefully intuitive) computations. 

The library contains \(32\) defined measurement systems:~\texttt{SI} which includes \(22\) basic and derived units; and \texttt{BIS} which includes \(5\) units for length (yards), mass (pound), temperature (rankine), volume (\(yard^3\)), and velocity (\(\frac{yard}{s}\)). Also, definitions of the \texttt{CGS} (Centimetre Gram Second) measurement system of units for length (centimeter), and mass (gram); and the \texttt{MHC} (Milligram Hour Celsius) measurement system of units for mass (miligrams), time (hour), and temperature (celsius). These allows for further derived units like foot, inches, square foot, day, week, year, \textit{etc}. 

\subsection*{\texttt{ISQ} use and other definitions}

We ported the most important theorems from Isabelle/HOL ISQ library~\cite{Physical_Quantities-AFP} and wrote them as VDM traces. We also added various expected equivalences between important measurement systems to hold as part of the library. These are defined through \(23\) VDM trace specifications, which lead to a total of \(2132\) trace tests run for nearly all defined measurement systems and their extension. This is useful to ensure computations are working as expected, and also to show users how to use the various library features in practice. 

We also define some \texttt{SI} common constants, such as the ceaseium frequency, speed of light, Planck constant, \textit{etc}. These also come from the Isabelle/HOL version of this library~\cite{Physical_Quantities-AFP}. Notice that non derived measurement systems are not defined as constants (\textit{e.g.}~\texttt{SI\_WATT}) but as constant functions (\textit{e.g.}~\texttt{SI\_JOULE()}). This is so that they are not computed at initialisation time.  

\subsection*{VDMJ high precision}

VDMJ provides a high precision version, which allows for unbounded precision on integers and higher precision on reals. The \texttt{ISQ} library is orthogonal:~it works the same with either version of VDMJ.\@ This means users will have to be careful with approximation within calculations.\@  

That is, when dealing with (potentially high precision) real-valued operations, it is important to consider approximations. We provide approximation functions for given orders of magnitude, as well as approximate equality module such order of magnitude (\textit{e.g.} \texttt{approx\_eq(pi, 3.14, 2) = true}). Another useful function we add that is not available in VDM is ceiling of a real value (\(\lceil r \rceil\)). 

\section{Applications and Examples}\label{sec:Examples}

This library was motivated after work various medical applications in a variety of (non standard) units alongside another variety of (familiar yet non-standard) time scales. For example, taking \(X\)mg of drug \(Y\) every \(8\) hours, which then has to be reviewed every \(4\) weeks for a year. Another concrete application was the use of different time abstractions having to be converted for understanding information within an embedded medical device for organ transplantation.  

\paragraph*{Dimension vector correspondence checking.}~
The library types enable users to document expected unit correspondences after an operation as type invariants. For example, the energy-pressure-volume (EPV) correspondence discussed above between \textit{Joule} (\(J\), amount of energy) and \textit{Pascal} (\(Pa: \frac{kg}{ms^2}\), amount of pressure) one might (mistakenly) write
\begin{vdmsl}[frame=none,basicstyle=\ttfamily\scriptsize]
    PA = ms_div(KILOGRAM, SI_ACCELERATION);
    p is_Pressure(PA) 
    = false

    PA' = ms_div(KILOGRAM, ms_times(METER, ms_itself_n(SECOND, 2)));
    p is_Pressure(PA')
    = true

    EPV = ms_times(PA', SI_VOLUME);
    p is_Energy(EPV)
    = true
\end{vdmsl}
\noindent That is, considering mass over acceleration (\(\frac{m}{s^2}\)) instead of mass dispersion over time. This also illustrates the use of measurement systems multiplication, replication and division. Finally, we check that the \textit{Pascal} measurement system alongside some volume equates to the expected unit of energy. These invariant checks (failures and successes) are useful to ensure that operations are being performed in the expected/adequate measurement system.  


\paragraph*{Dimension vector inspection.}~
Given that such computations over unit correspondences can test one's memory of physics (or other topics), the library also provides some useful utility functions to quickly inspect the dimension vector as one would use in ``pen-and-paper'' exercises. For instance, the dimension vector for the variables above can be viewed as
\begin{vdmsl}[frame=none,basicstyle=\ttfamily\scriptsize]
    p si_dim_view(PA)
    = "kg * (s**2) / m" 

    p si_dim_view(PA')
    = "kg / m * (s**2)"

    p si_dim_view(EPV)
    = "kg * (m**2) / (s**2)"
\end{vdmsl}   
\noindent This can be useful for debugging whether the unit of the measurement system chosen is the one expected. Notice that \texttt{si\_dim\_view} accepts a \texttt{Prefix}, hence making it seemless whether you give it a quantity or measurement system. 

Finally, the general \texttt{dim\_view} function expects an input that maps dimensions to their visualisation strings. For example, we add such a map to change the visualisation into \texttt{BIS}. This is not a check that the dimension is ``correct'' with respect to the measurement system conversion schema, but simply a debugging aid.     

\paragraph*{New measurement systems.}~
We created a new measurement system and conversion schema to cater for the  \texttt{MHC} measurement system. It is justs like the \texttt{SI} measurement system, yet views mass, time and temperature with these other corresponding units.
\begin{vdmsl}[frame=none,basicstyle=\ttfamily\scriptsize]
    MHC: ConversionSchema = CONV_ID ++ 
        { <Mass> |-> mag(milli(milli(KILOGRAM))), <Temperature> |-> -272.15,
          <Time> |-> mag(hour(SI, SI_UNIT)) };

    SECOND: Second = mk_MeasurementSystem(UNIT_TIME, SI, SI_UNIT);

    second: ConversionSchema * UnitSystem -> Second
    second(cs, u) == mk_MeasurementSystem(UNIT_TIME, cs, u);
    
    hour: ConversionSchema * UnitSystem -> Second 
    hour(cs, u) == scaleMS(60, scaleMS(60, second(cs, u)));      
\end{vdmsl}
\noindent The auxiliary functions create a conversion-schema parameterised notion of base unit of time (\textit{e.g.}~\texttt{SI}'s \texttt{SECOND} vs.\@ \texttt{second} possibly with different schema). In fact, even though our desired time baseline was hour, we added other granularities like minute, day, week, year, \textit{etc.} When evaluated the \texttt{MHC} system has \(\frac{1}{10^6}\) value for mass, \(3600\) for time, and \(-272.15\) for temperature. That is, usual mass (kilogram) is viewed in miligrams (mg), usual time (in seconds) is viewed in hours, and usual temperature (in kelvin) is viewed in centigrade.

Next, we created application-level types for each dimension within the measurement system, alongside corresponding single-unit constants for each type.
\begin{vdmsl}[frame=none,basicstyle=\ttfamily\scriptsize]
    Milligram = MHC_MeasurementSystem inv ms == ms.quantity.dim = MASS;	
    Hour      = MHC_MeasurementSystem inv ms == ms.quantity.dim = TIME;	
    Celcius   = MHC_MeasurementSystem inv ms == ms.quantity.dim = TEMP;

    MGRAM   : Milligram = mk_MeasurementSystem(UNIT_MASS, MHC, MHC_UNIT); 
    MHOUR   : Hour 	    = mk_MeasurementSystem(UNIT_TIME, MHC, MHC_UNIT);
    MCELCIUS: Celcius   = mk_MeasurementSystem(UNIT_TEMP, MHC, MHC_UNIT);
\end{vdmsl}	
\noindent These types and baseline constants were then refined in the actual application(s) to specific constraints like maximum treatment time or temperature. Then, we defined constant functions to convert other used time granularities back to the baseline in hours\footnote{We omitted the same type signature of other functions.}.
\begin{vdmsl}[frame=none,basicstyle=\ttfamily\scriptsize]
    hDAY: () -> Hour
    hDAY()  == scaleMS(HOURS_PER_DAY, MHOUR);
    hWEEK() == scaleMS(DAYS_PER_WEEK, hDAY());
    hYEAR() == scaleMS(DAYS_PER_YEAR, hDAY());

    hMONTH: WhichMonth -> Hour 
    hMONTH(m) == scaleMS(DAYS_PER_MONTH(m), hDAY());
\end{vdmsl}	
\noindent For a view of specific months in hours, we define a map from month to number of days and scale the result with \texttt{hDAY()}, hence creating month-specific hour periods. For the application-level we could have written something like (to hide away any ISQ details):
\begin{vdmsl}[frame=none,basicstyle=\ttfamily\scriptsize]
    n_times_day2every_x_hours: nat1 -> real
    n_times_day2every_x_hours(times_a_day) == 
        mag(scaleMS(1/mag(scaleMS(times_a_day, MHOUR)), hDAY()));
\end{vdmsl}
\noindent This function converts times per day in number of hours. Of course, this simple concrete example is somewhat overkill. It serves to show how the library can be used. 

For the real application, however, this was crucial, given variations were defined per week, where intake was not always per hour (\textit{e.g.}~three times a day), and other more complicated pharmacokinetic interactions existed.   

Other examples of use are provided in the \texttt{ISQ.vdmsl} distribution\footnote{\url{https://github.com/leouk/VDM_Toolkit/}}. They show various ISQ measurement systems in practical use.

\section{Results and discussion}\label{sec:Results}

ISQ representation and operations are an important part of modelling tasks within computing, physics and engineering. Unit conversion, combination and manipulation are notoriously error prone, having led to a number of serious accidents\footnote{\url{https://www.simscale.com/blog/nasa-mars-climate-orbiter-metric/}}.

The \texttt{ISQ.vdmsl} library closely follows the structure of a similar library for the Isabelle/HOL theorem prover~\cite{Physical_Quantities-AFP}. It has been used in a few practical industrial applications involving emedded systems and medical devices and processes.  

In this paper, we presented a formally defined ISQ library in VDM that adheres to our \textbf{SAFE} design principles (in~\Cref{sec:principles}). The library architecture (\Cref{sec:architecture}) is \textbf{S}imple, given its layered access to functionality. It is also \textbf{A}ccurate, given the presence of multiple kinds of user-defined invariant and other structural and data validation checks, such as enforcement of specific measurement system characteristics for well known quantities and units. It is \textbf{F}ast, as the computations are simple and directly available in VDM itself. Finally, it is \textbf{E}ffective, given its combination of speed, ease of use, multiple capabilities around ISQ measurement systems creation, combination, conversion, scaling, and so on.   

\paragraph*{Future work.}~
We are interested into applying the library concepts to other industrial applications, such as the Functional Mock-up Interface (FMI), which do have such notions for ISQ units\footnote{\url{https://fmi-standard.org/docs/3.0/\#_physical_units}}. Moreover, we also plan to extend the physical units to include all those in the SI tables\footnote{\url{https://en.wikipedia.org/wiki/International_System_of_Units}} and other more exotic list of units\footnote{\url{https://en.wikipedia.org/wiki/List_of_physical_quantities}}. 

\paragraph*{Acknowledgements.}~
We acknowledge funding from NCSC on verification on payment systems. We very much appreciate fruitiful discussion with and suggestions by Nick Battle about the usefulness of such a library and on certain VDMJ needs. 

\bibliographystyle{splncs03}
\bibliography{isq.bib}


\end{document}